\documentclass[figures]{epl}
\usepackage{amsmath}

\title{System size resonance in an attractor neural network}
\author{M. A. de la Casa\inst{1}\thanks{E-mail: \email{macasa@fisfun.uned.es}}
  \and E. Korutcheva\inst{1}{\bf \thanks{also G.Nadjakov Inst.Solid State
  Physics, Bulgarian Academy of Sciences, 1784 Sofia, Bulgaria}} \and J. M. R. Parrondo\inst{2} \and
  F. J. de la Rubia\inst{1}}
\shortauthor{M. A. de la Casa \etal}
\institute{
  \inst{1} Dpto. F\'{\i}sica Fundamental, Universidad Nacional de
    Educaci\'on a Distancia, c/Senda del Rey 9, 28040 Madrid, Spain \\
  \inst{2} Grupo Interdisciplinar de Sistemas Complejos {\em (GISC)} and Dep. F\'{\i}sica At\'omica,
  Molecular y Nuclear, Universidad Complutense, 28040 Madrid,
  Spain}

\begin{document}

\maketitle

\begin{abstract}
We study the response of an attractor neural network, in the ferromagnetic
phase, to an external, time-dependent stimulus, which drives the system
periodically toward two different attractors.  We demonstrate a non-trivial
dependence of the response of the system via a system-size resonance, by
showing a signal amplification maximum at a certain finite size.
\end{abstract}

\section{Introduction}

The counter-intuitive role of fluctuations as a source of order
has attracted much attention for the last
years~\cite{lefever,jordi}. Particularly interesting is the
phenomenon of stochastic resonance where the response of a
non-linear system to the action of a weak signal is enhanced, not
hindered, by the addition of an optimal amount of
noise~\cite{review1}. Among several potential applications of
stochastic resonance, there is evidence that it plays an important
role in some cognitive processes, such as perception
\cite{reshum,moos,reshum1}.

Another phenomenon that recently appeared in the literature and is
closely related to stochastic resonance is system size resonance,
or SSR from now on~\cite{yo}. In SSR, the presence of noise in a
system of finite size, close to a second-order phase transition,
gives rise to the appearance of an optimal size for the system to
adapt to an external field~\cite{yo,raul,hanggi}.

Inspired by the applications of stochastic resonance to cognitive
processes, in this Letter we show that SSR can operate in a simple
model of associative memory, namely, a Hopfield neural
network~\cite{Hopfield}, improving its ability to follow a
time-dependent stimulus. We will focus on the simplest case of a
Hopfield network storing just two patterns.

\section{Model}

The model is defined by the following Hamiltonian~\cite{Hopfield}:
\begin{equation}
\label{hamiltonian} {\cal H}=-\frac{1}{N} \sum_{i<j} J_{ij} s_i
s_j  -h\sum_i \xi_i^{\mu(t)} s_i ,
\end{equation}
where $s_i=\pm 1, i=1,...,N$ are  $N$-binary neurons and
$\xi_i^{\mu}=\pm 1,i= ,\dots,N, \mu=1,2$ are the two binary
patterns the system is trained with. We assume that the network
has been trained following the Hebb rule~\cite{Hebb}:
\begin{equation}
\label{coupling} J_{ij}= \xi_i^{1} \xi_j^{1}+\xi_i^{2} \xi_j^{2} .
\end{equation}

The second term in Eq.~(\ref{hamiltonian}) represents a non
constant but periodic stimulus of period $2T$: $\mu(t)=1$ if $t\in
[0,T)$ and $\mu(t)=2$ if $t\in [T,2T)$, which biases the system
towards pattern 1 and 2, alternatively. The intensity of the
stimulus is given by $h$.

A relevant magnitude in our analysis is the Hamming distance
between the patterns:
\begin{equation}
\label{distance}
d=\frac{1}{2N} \sum_{i=1}^{N} |\xi_i^{1}- \xi_i^{2}|,
\end{equation}
i.e., the fraction of sites in which both patterns are different.
It varies between $0$, when both patterns are equal at every site,
and $1$ when they are completely different.

This system has been extensively studied in the
literature, mainly in the thermodynamic limit $N
\rightarrow \infty$ (see~\cite{Eliza,Hertz} and the references therein). 
The role of finite size has only been
studied in order to find finite-size corrections to infinite-size
results~\cite{Privman}. In this work, we  keep $N$ explicitly
finite and focus on the effect of the finite size on the response
of the system to the time-dependent stimulus.

\section{Equilibrium states}

The two following order parameters are usually defined to describe
the macrostate of the network:
\begin{equation}
\label{ops} m_{\mu}=\frac{1}{N}\sum_i \xi_i^{\mu} s_i,
\end{equation}
which are the overlaps between each pattern and the neurons.

\begin{figure}
\onefigure[width=5cm]{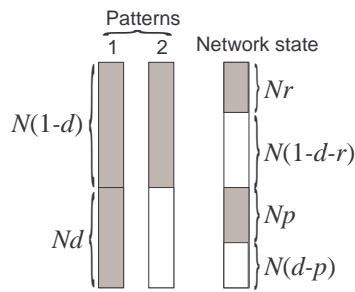} \caption {Schematic representation of the
two patterns and the state of the network.  The upper part of the patterns
represents the $N(1-d)$ coincident bits between the patterns. $Nr$ is the
number of bits that the network has in common with both patterns, whereas $p$
is number of bits in the network coinciding with pattern 1 and different from
pattern 2.}
\label{fig1}
\end{figure}

\begin{figure}
\twoimages[width=6cm]{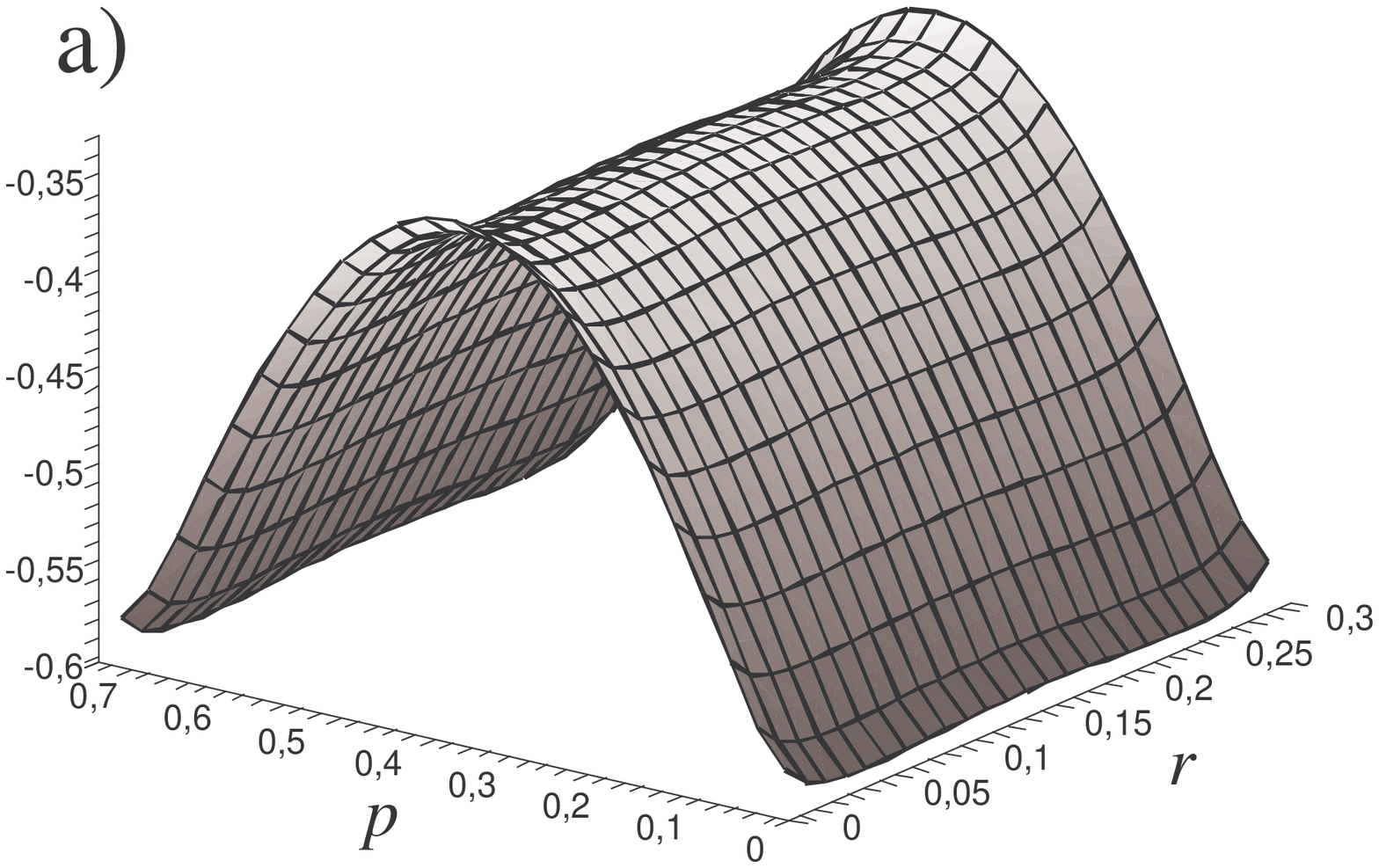}{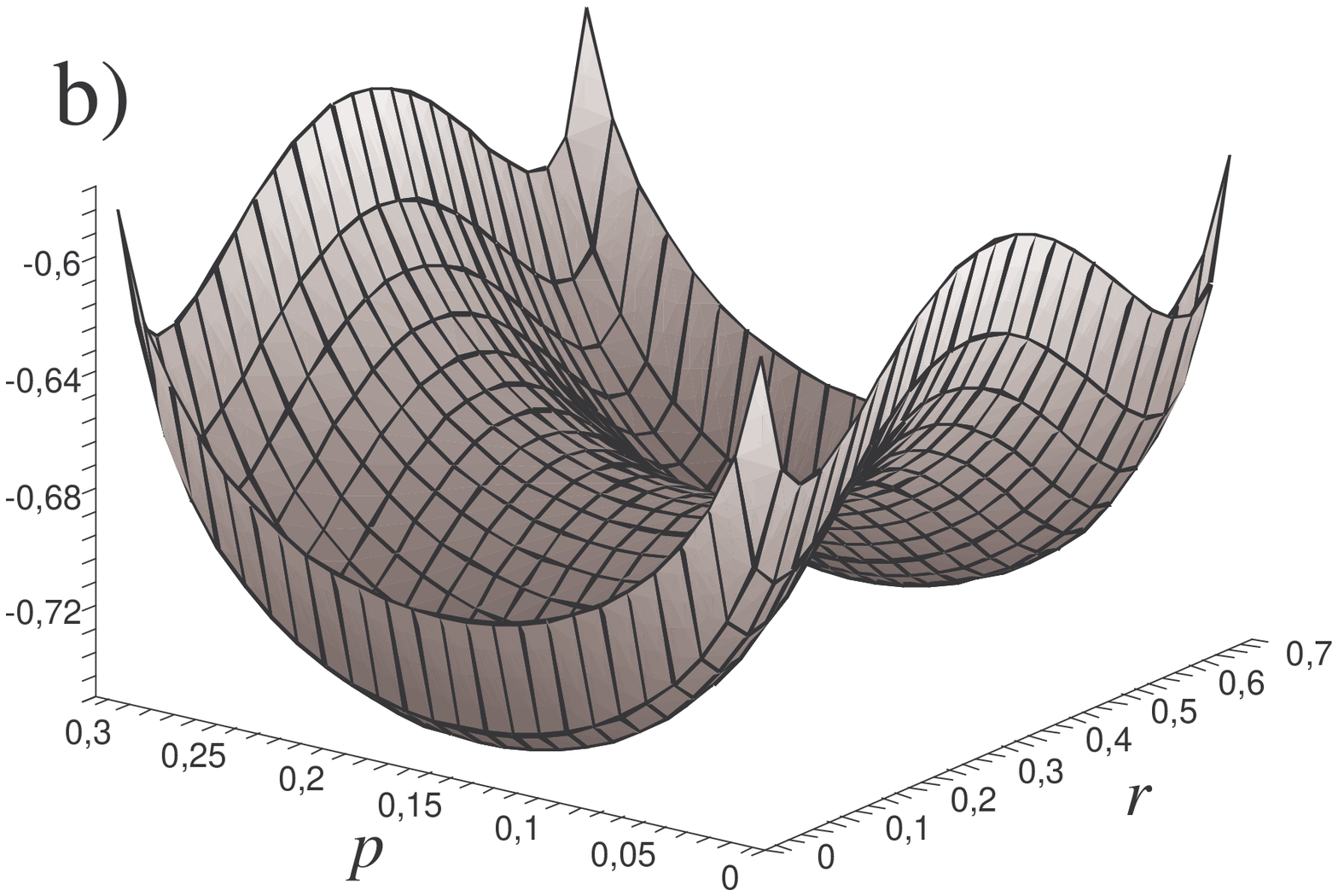}
\twoimages[width=6cm]{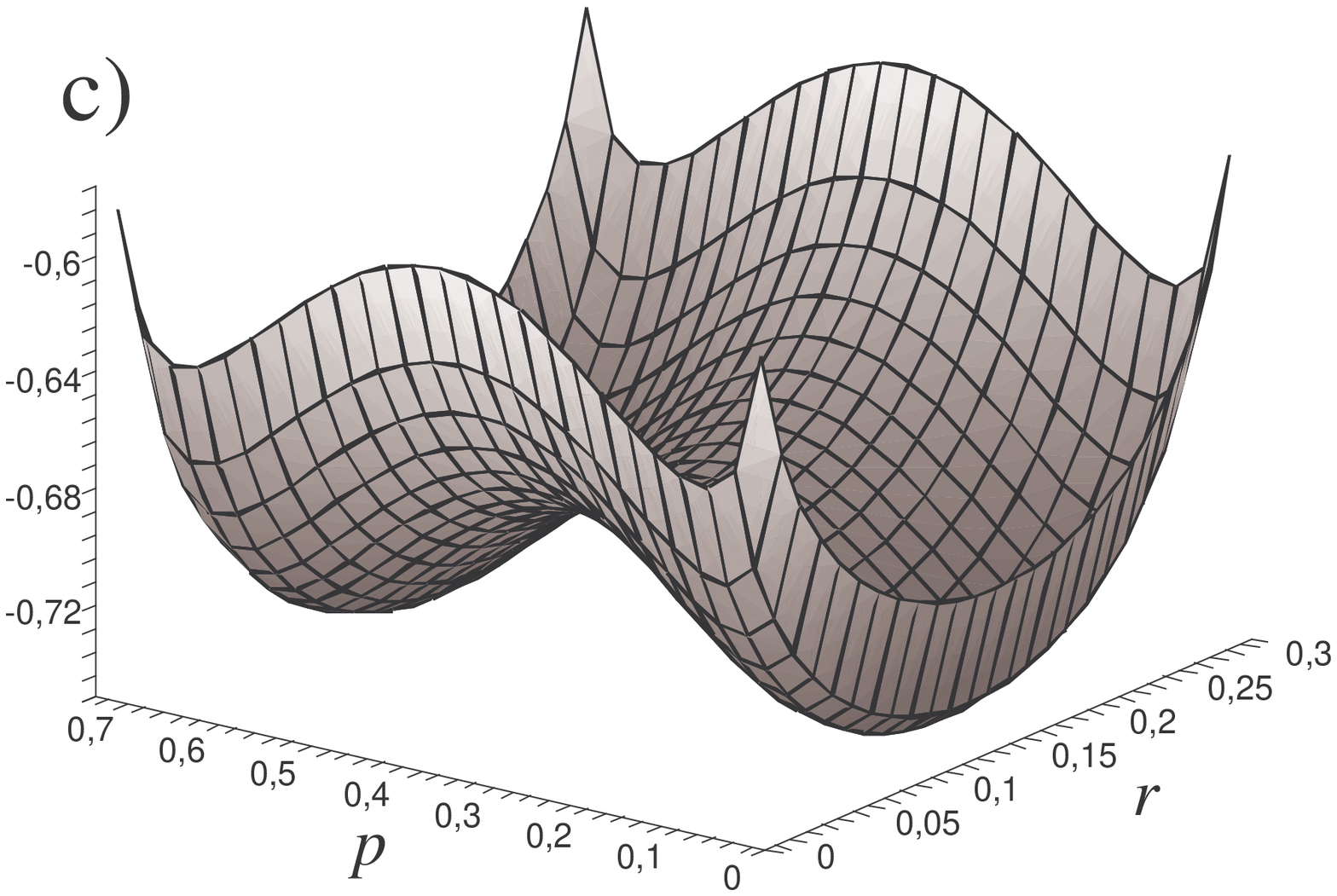}{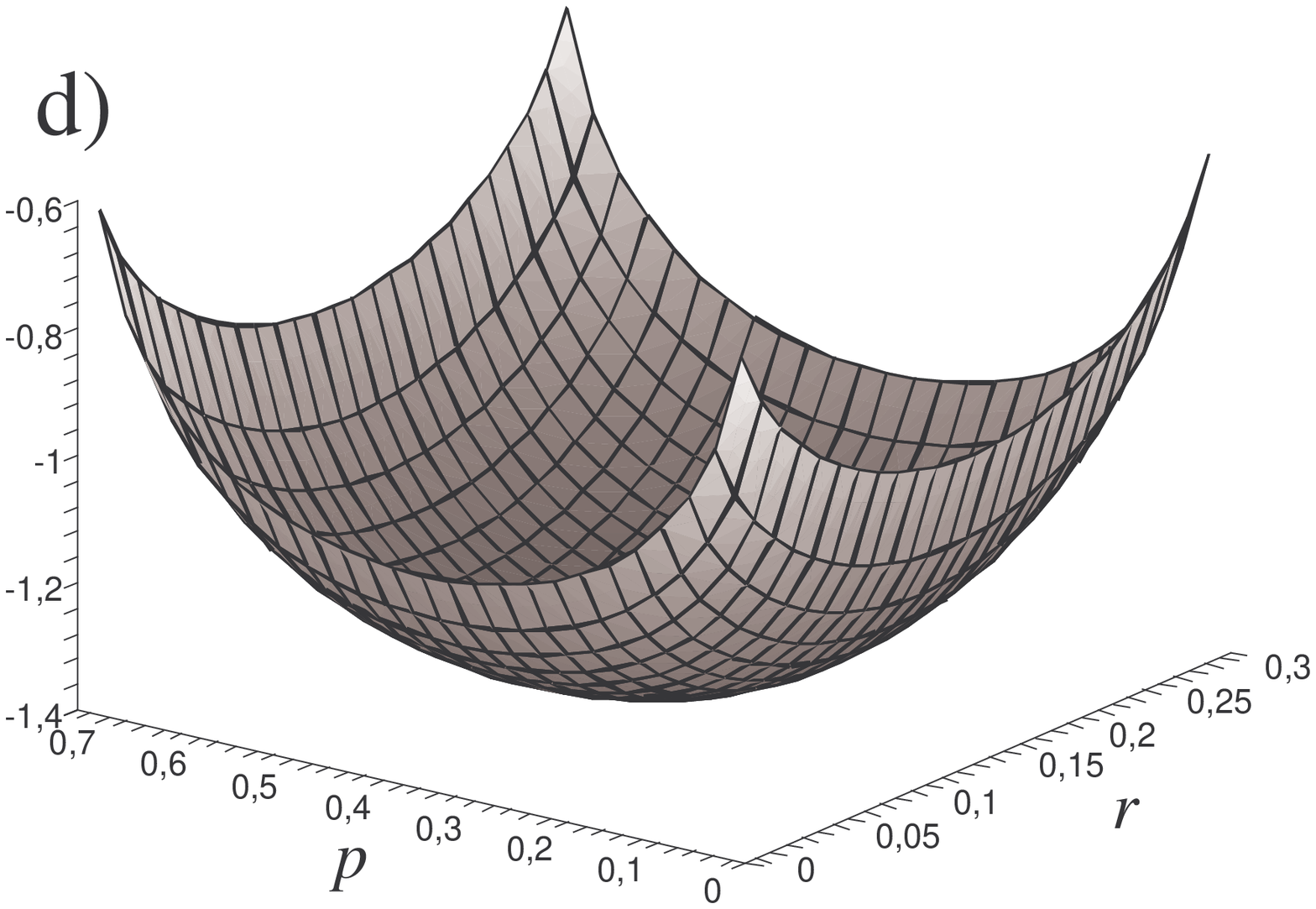} \caption { The free
energy landscape for $h=0$ and $N=10000$. a) {\em Low
temperature}, $\beta=2, d=0.7$: The free energy presents four
minima, corresponding to the two stored patterns and their
respective negatives.  b) {\em Medium temperature, similar
patterns}, $\beta=1, d=0.3$: There are two equilibrium states with
$p=d/2$, i.e., with $m_1=m_2$. One of the minima reproduces the
common bits of the two patterns whereas the other one is its
negative. c) {\em Medium temperature, dissimilar patterns},
$\beta=1$, $d=0.7$: There are two equilibrium states with
$r=(1-d)/2$, i.e., with $m_1=-m_2$. Each minima reproduces the
distinct bits of each pattern.  d) {\em High temperature},
$\beta=0.5$, $d=0.7$: The only minimum is the disordered state
with $r=(1-d)/2$, $p=d/2$, i.e., with $m_1=m_2=0$.} \label{phases}
\end{figure}

However, these two parameters are not independent. A more
convenient pair of magnitudes can be  defined as follows: $r$ is
the fraction of bits where $\{ s_i\}$ coincides both with pattern
1 and 2; $p$ is the fraction of bits where $\{ s_i\}$ coincides
with pattern 1 and differs from pattern 2. In Fig.~\ref{fig1} we
have plotted a schematic representation of the two patterns and
the quantities $p$ and $r$. Taking into account that $m_\mu$ is
the fraction of common bits minus the fraction of different bits
between the network and pattern $\xi^\mu$, from Fig.~\ref{fig1}
one immediately obtains:
\begin{equation} m_1 =2r+2p-1;\qquad m_2=2r+2d-2p-1.
\label{mpr}
\end{equation}
Also form this figure, it is easy to see that $r$ and $p$ are
independent and can take on any value in the rectangle: $r\in
[0,1-d]$, $p\in [0,d]$. Moreover, in terms of $r$ and $p$, the
free energy of the system for $h=0$ can be written as
\begin{equation}
{\cal F}={\cal F}_r+{\cal F}_p \label{free_energy}
\end{equation}
with \begin{eqnarray} {\cal F}_r &=&
-4N\left(r-\frac{1-d}{2}\right)^2-\frac{1}{\beta} \log \left(
\begin{array}{c}
N(1-d) \\
Nr
\end{array} \right) \nonumber \\
{\cal F}_p &=& -4N\left(p-\frac{d}{2}\right)^2 -\frac{1}{\beta}
\log \left(
\begin{array}{c}
Nd \\
Np\end{array} \right).
\end{eqnarray}
At zero temperature, the free energy has four minima, located at
the corners of the available rectangle in the parameter space,
i.e., $r=0,1-d$, and $p=0,d$. Two of the four equilibrium states,
$r=1-d$, $p=0,d$, exactly reproduce the two patterns. The other
two minima, $r=0$, $p=0,d$, are the negatives of the stored
patterns. Fig.~\ref{phases} a) shows a typical landscape of the
free energy in the $(r,p)$ plane for low temperatures ($\beta=2$,
$d=0.7$).

If we increase the temperature, the four minima shift to the
middle point of the rectangle $r=(1-d)/2$ and $p=d/2$. In the
thermodynamic limit, the system undergoes two second-order phase
transitions at $\beta_{c,r} =1/(2-2d)$ and $\beta_{c,p}=1/(2d)$.
In each transition the minima collide into either $r=(1-d)/2$ or
$p=d/2$, which corresponds, respectively, to a completely
disordered state in the region of common and distinct bits between
the two patterns (see Fig.~\ref{fig1}). The ability of the network
to distinguish between the two patterns sensibly depends on which
of the two transitions occurs first.

If $d<0.5$, i.e., if the two patterns share more than a half of
the bits, $\beta_{c,p}>\beta_{c,r}$. Consequently, when the
temperature increases from absolute zero the first transition
occurs for the variable $p$, i.e., in the region of distinct bits.
This means that for temperatures $\beta\in
[\beta_{c,r},\beta_{c,p}]$, the system only exhibits two minima
with $p=d/2$, i.e., with $m_1=m_2$ [see Eq.~(\ref{mpr})]. One of
these two minima approximately reproduces the common bits of the
two patterns, whereas the distinct bits are completely disordered.
The other minima is just the negative image of the first.
Consequently, the system has mixed up the two patterns and it is
unable to distinguish between them. The free energy landscape
corresponding to this situation is plotted in Fig.~\ref{phases}
b).

On the other hand, if $d>0,5$, the two patterns are different
enough to be distinguished even for intermediate temperatures. In
this case $\beta_{c,r}>\beta_{c,p}$, and the first transition
occurs at $\beta_{c,r}$. Therefore, if $\beta\in
[\beta_{c,p},\beta_{c,r}]$, we  have two minima with $r=(1-d)/2$,
i.e., with $m_1=-m_2$. One of the two minima reproduces the
distinct bits of pattern 1 and the other one the distinct bits of
pattern 2. For both minima, the common bits are disordered.
Although the system does not exactly reproduce the stored
patterns, it perfectly distinguish between them. The free energy
in this case is plotted in Fig.~\ref{phases} c).

Finally, above the maximum critical temperature, the only
equilibrium state is completely disordered: $r=(1-d)/2$ and
$p=d/2$, or $m_1=m_2=0$, as shown in Fig.~\ref{phases} d).

Along this Letter we will focus only on the third case: patterns
with $d>0.5$ and temperatures corresponding to the landscape in
Fig.~\ref{phases} c). The reason is that the system still
distinguish between the two patterns, but we can reach
temperatures high enough to clearly observe SSR.
\section{Results}
We have performed out-of-equilibrium Monte Carlo simulations~\cite{newman} at
inverse temperature $\beta$ using the Hamiltonian~(\ref{hamiltonian}). The
dynamics is defined by a standard Metropolis algorithm in which
100000 sweeps have been carried out averaging over 100 realizations
for every parameter set.

\begin{figure}
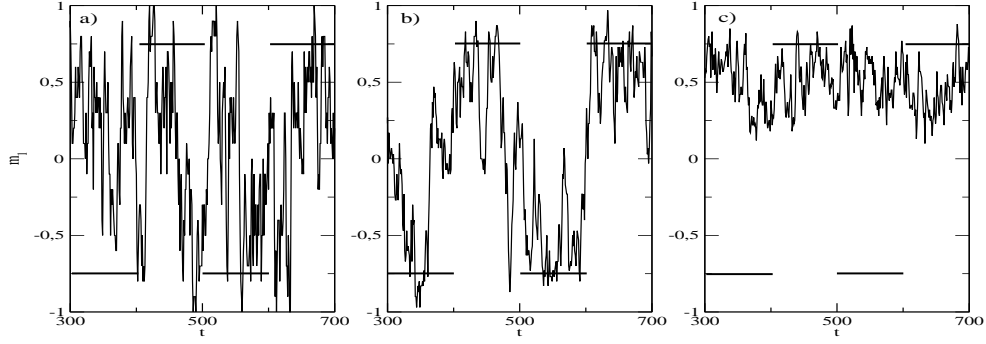

\onefigure[height=4.5cm,width=13cm]{m_vs_t.eps} \caption{Time
evolution of the order parameter $m_1$ showing the first evidence
of system-size resonance. The solid thin line is the order
parameter obtained from numerical simulations. The thick segments
show the time intervals in which the external stimulus drives the
system to retrieve one or the other pattern. The parameters values
are: $\beta=1.2$, $d=0.6$, $T=100$, $h=0.01$. a) $N=20$; b)
$N=60$; c) $N=120$.} \label{example}
\end{figure}

The simulations show a clear evidence of SSR. A relevant example
is presented in Fig.~\ref{example}, with the distinctive features
of stochastic resonance. For small $N$, fluctuations are strong
and the system output is too noisy being unable to retrieve the
patterns dynamically. The hoping between the attractors is random
and not synchronized with the switches of the external stimulus.
For very large $N$, fluctuations are weak and the system is
quenched in a given attractor. However, for intermediate values of
$N$, the system follows the oscillations of the external stimulus
and the appropriate pattern for every half-period is retrieved
very precisely.
In order to obtain a more quantitative picture of SSR, we use the
power spectrum $S_m(\omega)$ of one of the order parameters. We
will focus on the power spectrum of $m_1(t)$, but similar results
are obtained if $m_2(t)$ is chosen. A measure of the quality of
the response of the system to the external input is the so-called
{\em signal amplification} $\eta$~\cite{jung},
 defined as the ratio between the power spectrum
 at the external frequency $\Omega= \pi /T$ and
the total power contained in the external stimulus:
\begin{equation}
\label{amplification}
\eta=\lim_{\Delta \omega \rightarrow 0} \frac{2 \int^{\Omega+\Delta
\omega}_{\Omega-\Delta \omega} S_m(\omega) d \omega}{h^2}.
\end{equation}
Following previous work~\cite{yo,jung}, we will use the signal
amplification $\eta$ to locate and assess resonance phenomena.

We have also obtained an analytical expression for $S_m(\omega)$
reproducing quite well the numerical experiments. The main idea is
to approximate the dynamics of the network by a two-state system
and follow the calculation performed in~\cite{review1,mcnamara}.
The details of the theory will be presented in a forthcoming
publication~\cite{preparation}. Here we will only sketch the main
steps of the calculation.

The starting point is a master equation for the two-state system.
The states are the two minima of the free energy, which are
located numerically. As shown above, each of these minima
reproduces fairly well the two stored patterns, hence we label
them as 1 and 2. The transition probabilities between these two
states $W_{1\to 2}$ and $W_{2\to 1}$ depend on time as $W_{1\to 2}(\mu (t))$,
$\mu(t)$ being the pattern shown to the network at time $t$.
We have chosen an Arrhenius-like expression as in~\cite{review1,mcnamara}:
\begin{eqnarray}
\label{rate}
W_{1\to 2}(1) & = & c \exp (-\beta \Delta {\cal F}) \exp (- \beta h d_{12} N/2)
\nonumber \\
W_{1\to 2}(2) & = & c \exp (-\beta \Delta {\cal F}) \exp (\beta h d_{12} N/2),
\end{eqnarray}
where $d_{12}$ is the difference between the values of the order parameter
$m_1$ in the two minima, $\Delta {\cal F}$ is the free energy barrier
separating, at zero external field, the two minima along the path $r=(1-d)/2$,
with $m_1=-m_2$, and $c$ is an arbitrary constant (see panel c in
Fig.~\ref{phases}). The transition probabilities $W_{i\to j}$ are chosen to
satisfy detailed balance: $W_{1\to 2}(1)=W_{2\to 1}(2)$.

The corresponding master equation for this two-state stochastic
process can be solved exactly to get the time-dependent moments
$<m_1(t)>$ and $<m_1(t)m_1(t+\tau)>$. Finally, by using
Wiener-Khinchin theorem~\cite{gardiner}, we obtain the following
explicit expression for the power spectra $S_m(\omega)$:
\begin{eqnarray}
 {\cal S}_m (\omega)=\frac{d_{12}^2}{4} \left[
1-\tanh^2(\beta h d_{12} N) \left( 1-
  \frac{2\Omega}{\pi W} \tanh(\frac{\pi W}{2\Omega}) \right) \right]
  \frac{2 W}{\omega^2+ W^2} \nonumber \\
  + \frac{2 d_{12}^2}{\pi} \tanh^2(\beta h d_{12} N)
  \sum_{k=1}^{\infty} \frac{W^2}{(2k-1)^2 (W^2+(2k-1)^2 \Omega^2)}\nonumber  \\
  \times \left(\delta[\omega-(2k-1)\Omega]+\delta[\omega+(2k+1)\Omega]
  \right)
  \label{spectrum}
\end{eqnarray}
with
\begin{equation}
\label{W} W=W_{1\to 2}(1)+W_{1\to 2}(2)=c \exp (-\beta \Delta {\cal F}) \cosh(\beta h
d N).
\end{equation}
\begin{figure}[h]
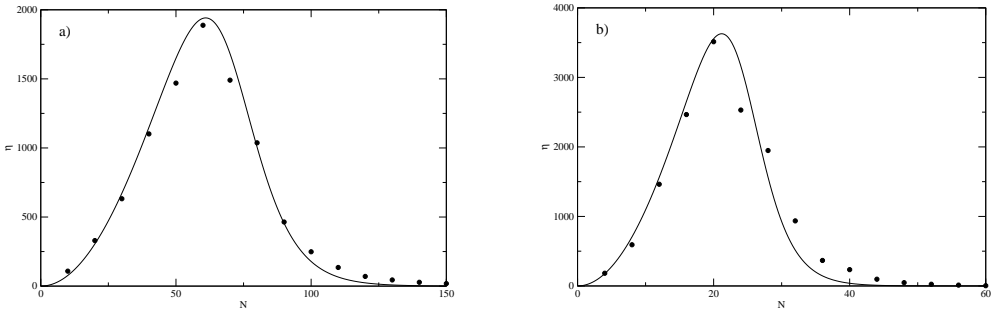

\twoimages[width=6cm]{s_vs_N_d06_T500_new.eps}{s_vs_N_d075_T500_new.eps}
\caption{The signal amplification, $\eta$, versus the size of the system, $N$
for $h=0.01$. The result of the numerical simulation is represented by dots
while the analytical result is the solid line. a) $d=0.6$, $T=500$,
$\beta=1.2$. b) $d=0.75$, $T=500$, $\beta=1.2$. For larger values of $d$, the
maximum is shifted to the left. The only free parameter in this theory is $c$,
which has been chosen to fit best the Monte Carlo data. In a), $c=0.44$; in
b), $c=0.65$.}
\label{vs_N}
\end{figure}
The resonant behavior of the signal amplification $\eta$ can be
seen in Fig.~\ref{vs_N}. There is a maximum of the signal
amplification at a finite size. The results of the numerical
simulation (dots) correspond very well to the analytical results
(solid line) given by Eq.~(\ref{spectrum}). The
difference between both panels of Fig.~\ref{vs_N} is exclusively
due to a difference in the value of the Hamming distance $d$,
keeping the other parameters $\Omega,h$ and $\beta$ equal. A
larger value of $d$ implies a higher energy barrier between the
attractors which needs  a larger noise intensity or smaller $N$ to
achieve the best resonance. Consequently,  the maximum of $\eta$
is shifted towards smaller $N$, and its maximum resonant value is
reduced.
\begin{figure}[h]
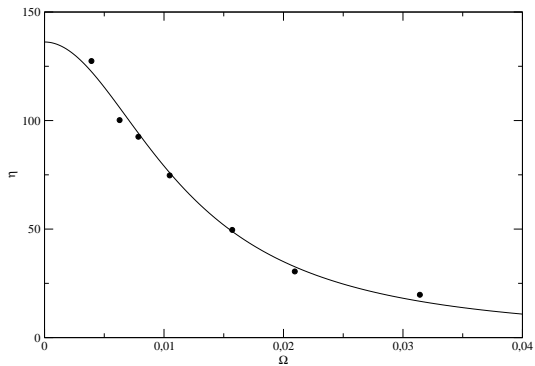

\onefigure[width=7cm]{s_vs_T_d06_N50_new.eps} \caption{The behavior of $\eta$
as a function of the frequency $\Omega$ of the external stimulus shows the
expected Lorentzian form, as in usual stochastic resonance. Again, the dots
represent the numerical simulations and the solid line the analytical
results. In this plot, $d=0.6$, $N=50$, $\beta=1.2$ and $h=0.01$. c=0.44, as
in panel a) in Fig.~\ref{vs_N}.} \label{vs_W}
\end{figure}
Finally the dependence of $\eta$ on the stimulus frequency
$\Omega$ is shown in Fig.~\ref{vs_W}. It shows the Lorentzian
dependence expected in SR~\cite{review1,review2}. A good agreement
between theory and numerical simulations is observed as well.
\section{Conclusions}
In this paper we have presented numerical simulations and
theoretical calculations, based on a two-state model, showing the
presence of system size resonance effects in an attractor neural
network. These effects are made evident by the resonant behavior
of the signal amplification $\eta$ as a function of the size of
the system, as well as by the time evolution of the order
parameters. We also point out the good agreement between
analytical results and simulations.

Our model shows that noise can provide the flexibility that an
adaptive memory needs to follow a time-dependent stimulus. Since
noise depends on the size of the system, we conclude that there is
an optimal size for which there is maximal synchronization of the
system to the evolving stimulus. We have explicitly shown this
resonant phenomenon in a simple model with two attractors, but it
is likely that more complicated models exhibit the same effect.

\acknowledgments This work is financially supported by Ministerio de Ciencia y
Tecnolog\'{\i}a (Spain), Projects No. BFM2001-291 and FIS2004-271, and by
UNED, Plan de Promoci\'on de la Investigaci\'on 2002.

\end{document}